# Idology and Its Applications in Public Security and Network Security

Shenghui Su [1, 4, 6(✉)], Jianhua Zheng [2, 6], Shuwang Lü [2, 5, 6], Zhiqiu Huang [1, 6], Zhoujun Li [3, 6], and Zhenmin Tang [4, 6]

[1] College of Computers, Nanjing Univ. of Aeronautics & Astronautics, Nanjing 211106, PRC
[2] Laboratory of Information Security, Univ. of Chinese Academy of Sciences, Beijing 100039, PRC
[3] School of Computers, Beihang University, Beijing 100191, PRC
[4] Center for Public Security Innovation, Nanjing Univ. of Science & Technology, Nanjing 210094, PRC
[5] Laboratory of Computational Complexity, BFID Corporation, Beijing 100098, PRC
[6] Preparatory Team for International Idology Research Association, Nanjing 211106, PRC

**Abstract.** Fraud (swindling money, property, or authority by fictionizing, counterfeiting, forging, or imitating things, or by feigning other persons privately) forms its threats against public security and network security. Anti-fraud is essentially the identification of a person or thing. In this paper, the authors first propose the concept of idology – a systematic and scientific study of identifications of persons and things, and give the definitions of a symmetric identity and an asymmetric identity. Discuss the converting symmetric identities (e.g., fingerprints) to asymmetric identities. Make a comparison between a symmetric identity and an asymmetric identity, and emphasize that symmetric identities cannot guard against inside jobs. Compare asymmetric RFIDs with BFIDs, and point out that a BFID is lightweight, economical, convenient, and environmentalistic, and more suitable for the anti-counterfeiting and source tracing of consumable merchandise such as foods, drugs, and cosmetics. The authors design the structure of a united verification platform for BFIDs and the composition of an identification system, and discuss the wide applications of BFIDs in public security and network security – antiterrorism and dynamic passwords for example.

## 1 Introduction

Fraud (swindling money, property, or authority by fictionizing, counterfeiting, forging, or imitating things, or by feigning other persons in private) exists ubiquitously in both the physical world and the digital world (cyberspace). It is age-old but active, and prohibited by laws but not prevented efficiently with techniques.

Fraud forms its threat against public securities. For instance, counterfeiting of merchandise brands (especially food and drug brands) [1][2], forgery of papery diplomas or certificates [3], faking of financial bills or notes [4], fiction of financial accounts [5], etc.

Fraud forms its threat against network securities. For instance, fiction of IP addresses, juggle of IP addresses [6], non-license of programs (note that the execution of malicious program is ascribed to non-license) [7], non-authorization of websites (namely fishing websites) [8], forgery of electronic documents [9], wiretapping of user passwords [10], etc.

Fraud is different from secret-outing. Keeping of secret relies on the application of encryption technology while prevention of fraud relies on the improvement and application of identification technology. Sometimes, symmetric ciphers are used for authentication, but it is not strict identification, and only the re-meeting between two friends who hold the same private key.

Throughout this paper, unless otherwise specified, the sign % means "modulo", $\bar{M}$ means "$M - 1$" with $M$ prime, lg $x$ denotes the logarithm of $x$ to the base 2, $\neg b_i$ does the opposite value of a bit $b_i$, Ꝑ does the maximal prime allowed in a coprime sequence, $|x|$ does the absolute value of a number $x$, $\|x\|$ does the order of an element $x$ % $M$, ¦$S$¦ does the size of a set $S$, and gcd($a$, $b$) represents the greatest common divisor of two integers. Without ambiguity, "% $M$" is usually omitted in expressions.

## 2 Several Definitions Relevant to Identification

In terms of american dictionaries, an identity is said to be *the set of characteristics* by which a person or thing is definitively recognizable or known, *the awareness* that an individual or group has of being a distinct and persisting entity, *the condition* of being a certain person or thing, *information* such as an identification number used to establish or prove a person's individuality [11], or *the set* of behavioral or personal characteristics by which an individual is recognizable as a member of a group [12]. Therefore, the identity concept is not only applicable to a person, but also a thing such as a material article, a machine, an organization, etc.

### 2.1 Identity, Subject, Object, and Idology

An identity referred here is not the social attribute of a person or thing — a nobleman or commoner, a luxury or pedlary for example, but the natural attribute of a person or thing, relating to inherence — the head portrait or fingerprint of a person stored in the chip of a passport for example.

***Definition 1***: An identity is a congenital mark of a person or thing (including organization) by which the nativity, derivation, affiliation, or / and distinctiveness of the person or thing are determined.

***Definition 2***: The administration of persons, the registry of users or organizations, the producer, maker, issuer, or approver of things is called a subject, and a person or thing is called an object.

Two affairs, namely the prevention of secret-outing and prevention of fraud, occurred simultaneously along with human beings wars, which indicates that identification technology is the same archaic as encryption technology. For example, in 257 B.C. belonging the Warring States Period of China, a son of the king of Wei State whose feoff was in XinLing rescued Zhao State successfully by stealing the half BingFu, dispatching the troop, reenforcing Handan, and defeating the enemy from Qin State [13].

A BingFu is a commander's tally which convinces the chief general at a military base that an order to maneuver troops comes indeed from the king or emperor of a state or nation, and is split into two halves — the right reserved by the king or emperor and the left given to the chief general (Fig. 1).

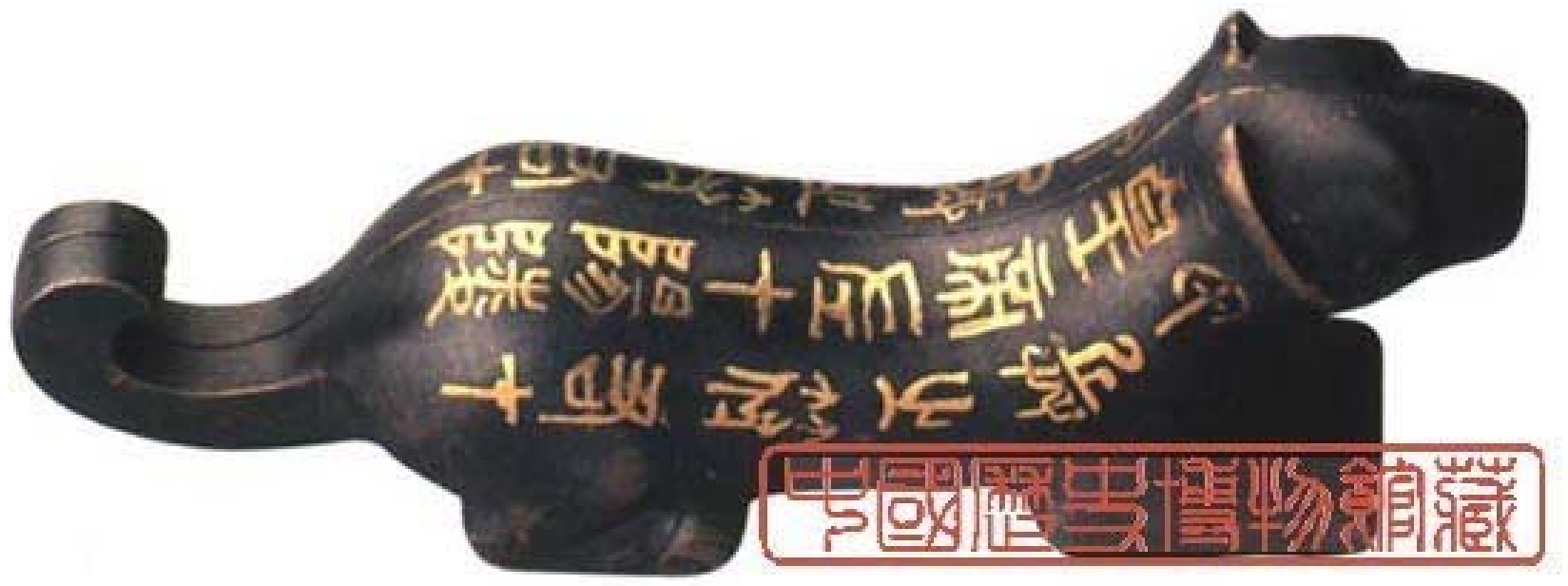

**Fig. 1.** A BingFu — the identity of military orders of the first emperor of the Qin Dynasty

In Fig. 1, the right and left of the BingFu have the same inscriptions: This is a tally for maneuvering soldiers in armour, the right is reserved by the emperor, and the left is kept by the chief general at YangLing.

Notice that the inscriptions of the right and left are arranged *asymmetrically*. When the first emperor maneuvers troops, the right must coincide with the left (Fig. 2).

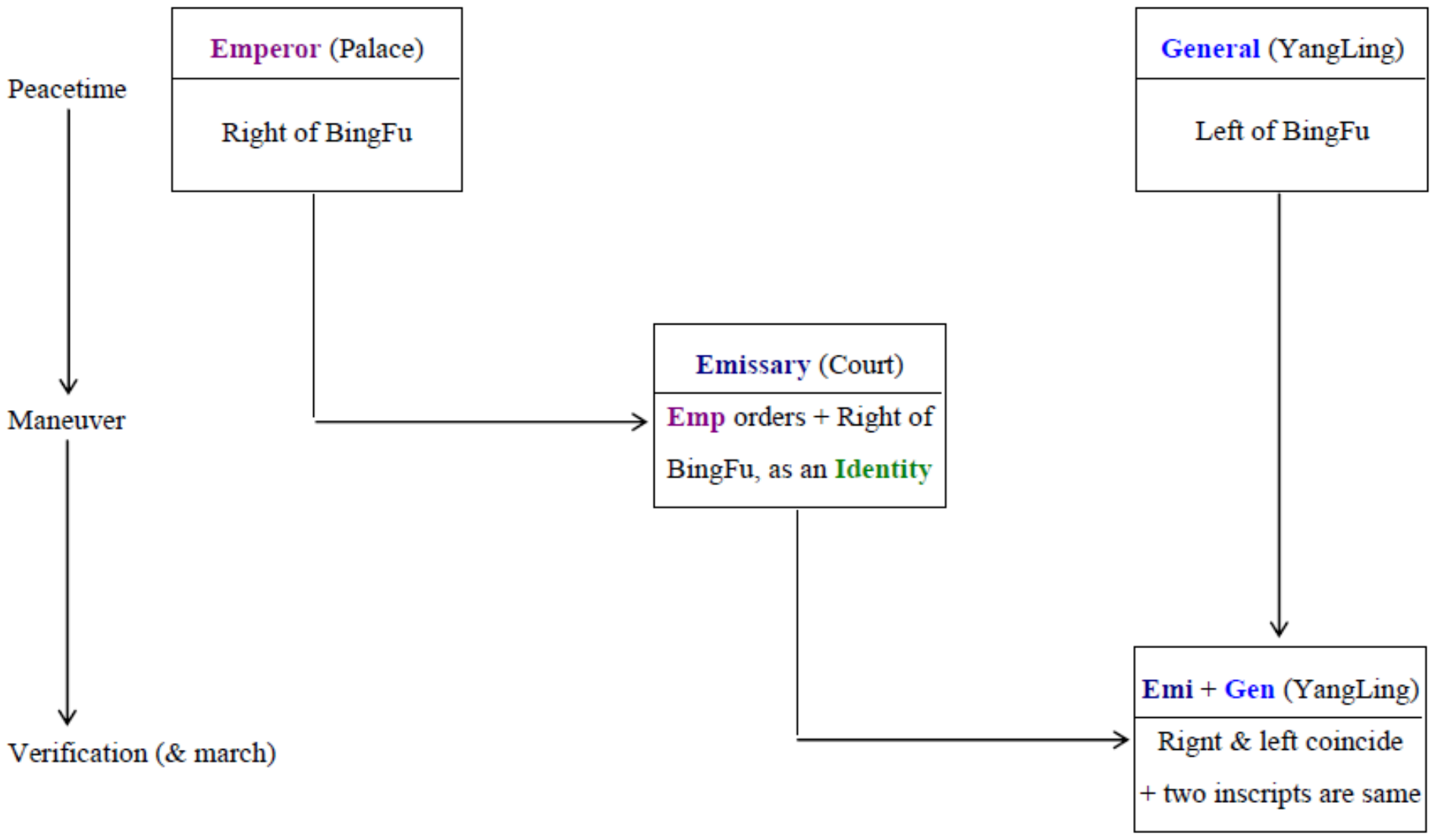

**Fig. 2.** Process of maneuvering troops with a BingFu

***Definition 3***: Idology is a systematic and scientific study of identity confection, identity denotation, identity sensing, and identity verification for the identification of a person or thing.

Identity confection is to say to make the characteristics of a subject and / or the characteristics of an object into an identity which is a string, graph, or cast.

Identity denotation is to say to set the identity of an object into a place which is a part of the object, or bound with the object.

Identity sensing is to say to detect (or perceive) an identity manually or automatically, extract the identity essence, and send (or transmit) it to a verification platform.

Identity verification is to say to affirm the correctness or facticity of an identity by evidence and logic, and on occasion return the result and related information to an inquirer [14].

As a branch of knowledge or teaching, idology has its research purpose, theoretical foundation, and exploration field.

The research purpose of idology is to prevent fraud in both the physical world and the digital world.

The theoretical foundation of idology involves computational complexity, informatics, number theory, abstract algebra, the design and analysis of algorithms, data structure, hash function, digital signature, security proof theory, etc.

The exploration field of idology includes symmetric identities, asymmetric identities, isomeric identities, communication networks, internet of things, big data, cloud computing, pattern recognition, biometrics, electronic or imaging sensors, software engineering, etc.

## 2.2 Symmetric Identity and Asymmetric Identity

On the view of practice, identities present symmetric and asymmetric modalities.

### 2.2.1 Symmetric Identities

***Definition 4***: A graph or a string which is verified through direct comparison or simply computational comparison with a duplicate stored in advance is called a symmetric identity.

The evolution of symmetric identities experiences two phases.

◈ Classical Symmetric Identities

They include passwords or watchwords which emerged first within the ancient Roman military (Fig. 3) [15], stamps (equivalent to handwritten signatures) which occurred at least two thousand years ago — the nephrite stamp of the empress in the Western Han Dynasty for example (Fig. 4) [16], watermarks of which the idea was first brought forward by an Italian in 1282 (Left of Fig. 5) [17], trademarks which are thought of as being used first by blacksmiths of the Roman Empire period from27 BC to 476 AD (Right of Fig. 5) [18], etc.

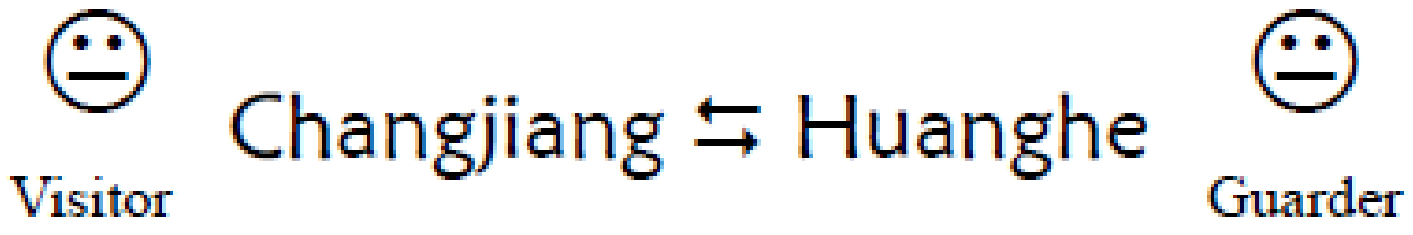

**Fig. 3.** A password and a response to it

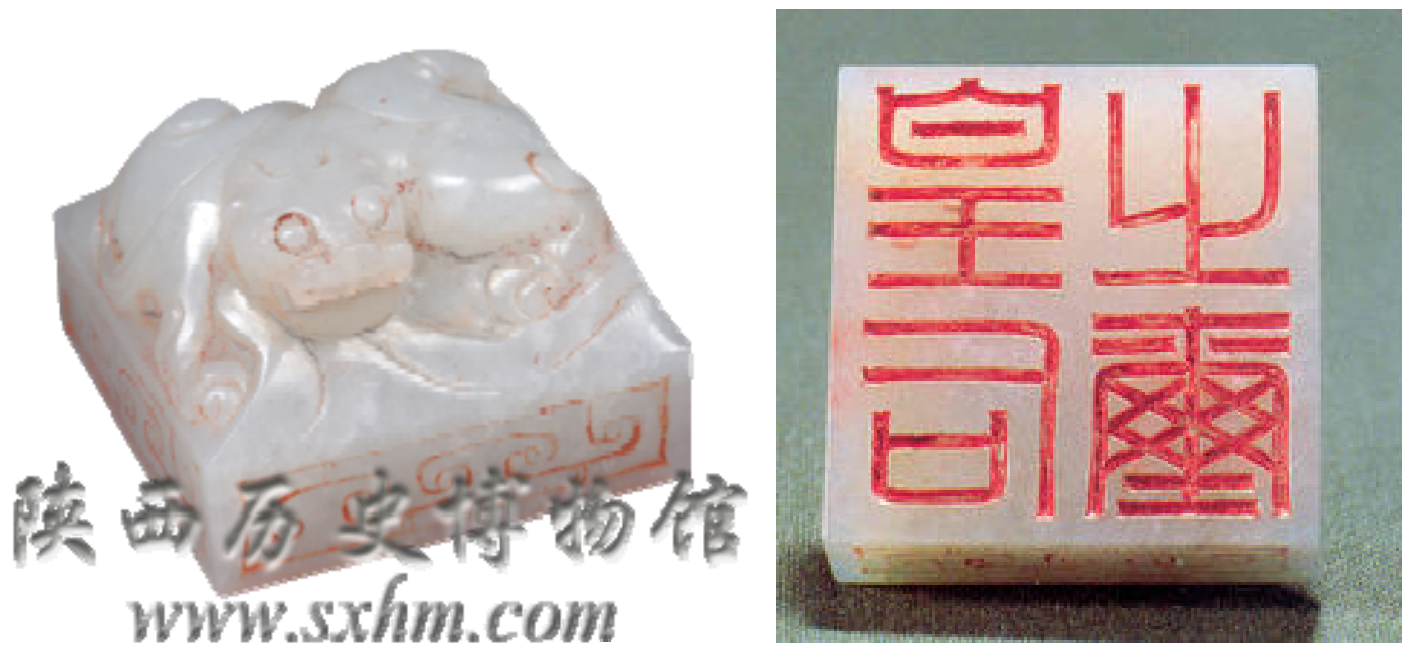

**Fig. 4.** The nephrite stamp of the empress

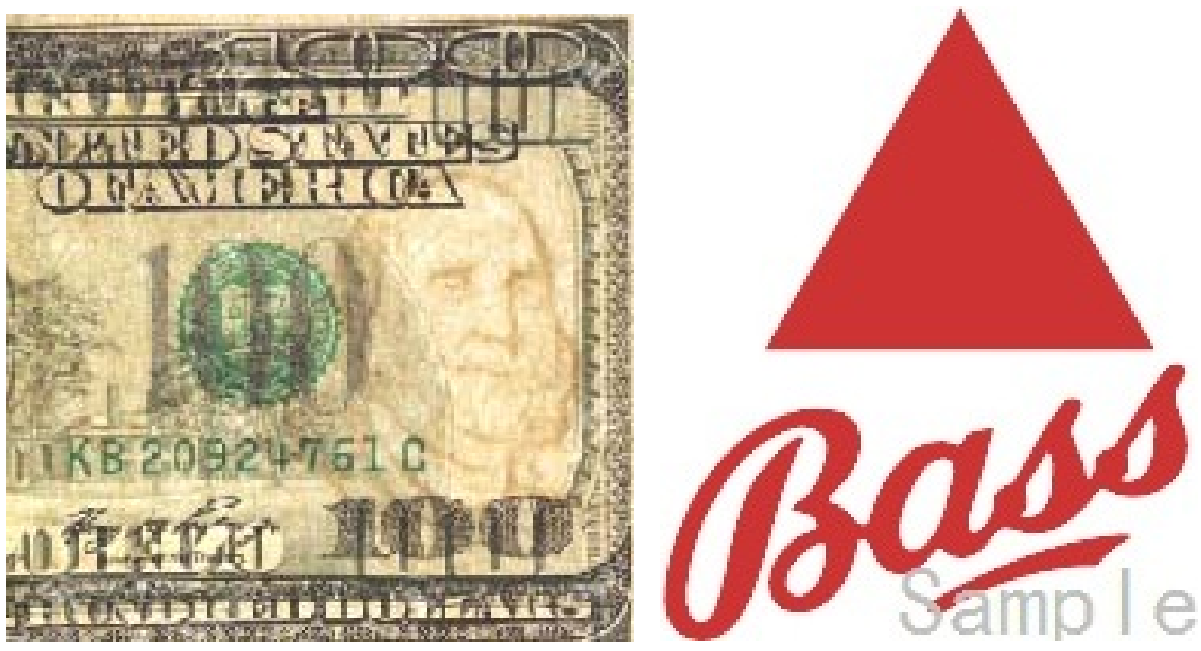

**Fig. 5.** A watermark and trademark

◈ Modern Symmetric Identities

They include holographic labels which occurred in the late 1970's (Left of Fig. 6) [19], electronic query / supervision codes which emerged in the early 1990's (Fig. 7) [20], quick response codes (namely two-dimensional codes) which occurred in 1994 (Right of Fig. 6) [21], etc.

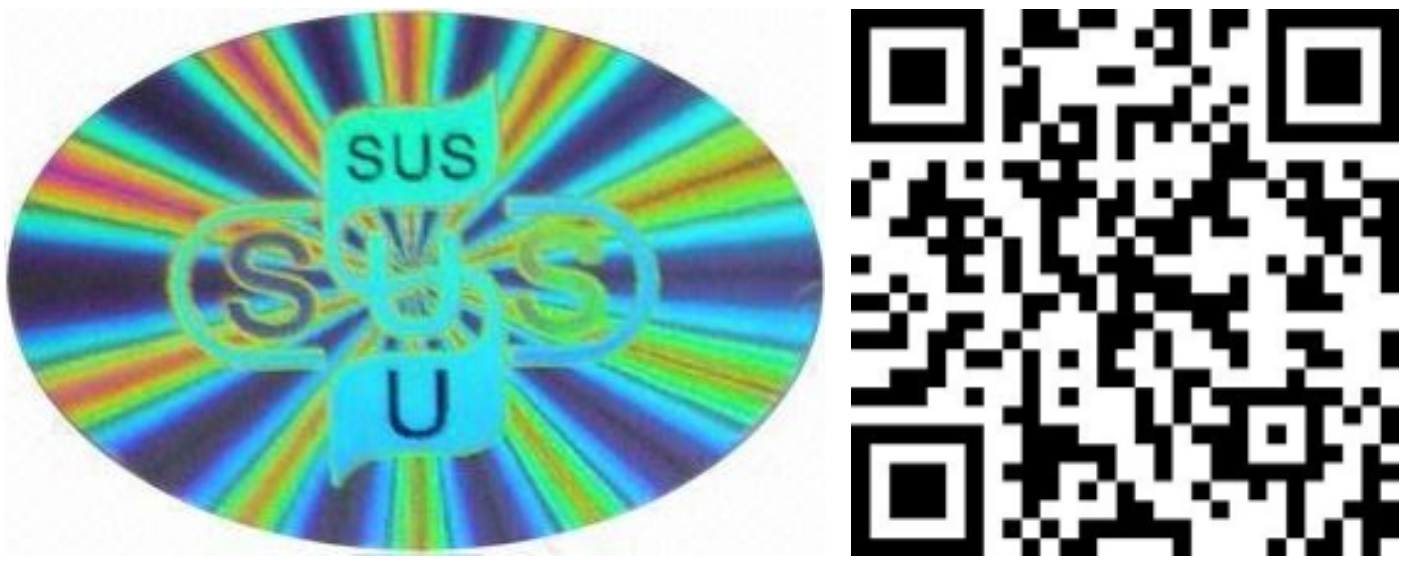

**Fig. 6.** A holographic label and QR code

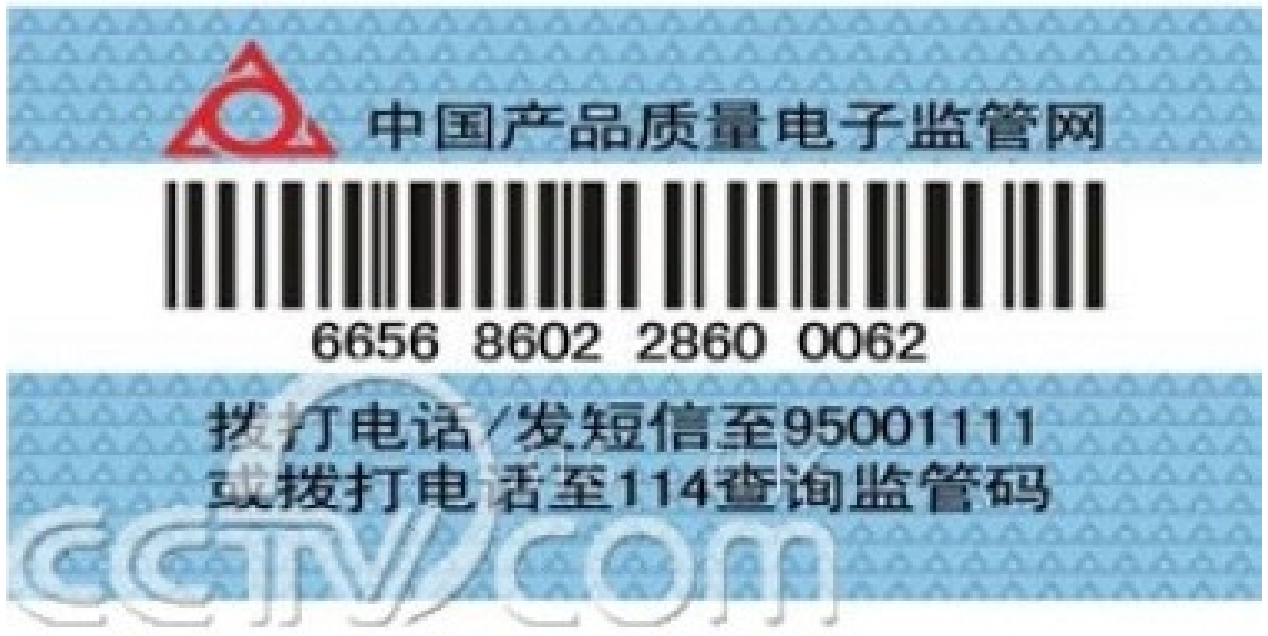

**Fig. 7.** An electronic supervision code

Especially, it should be pointed out that human biological characteristics such as a fingerprint and iris (Fig. 8) [22] are also a type of symmetric identity, where a fingerprint was applied to biometric authentication and criminalistics in the early 20th century [23][24].

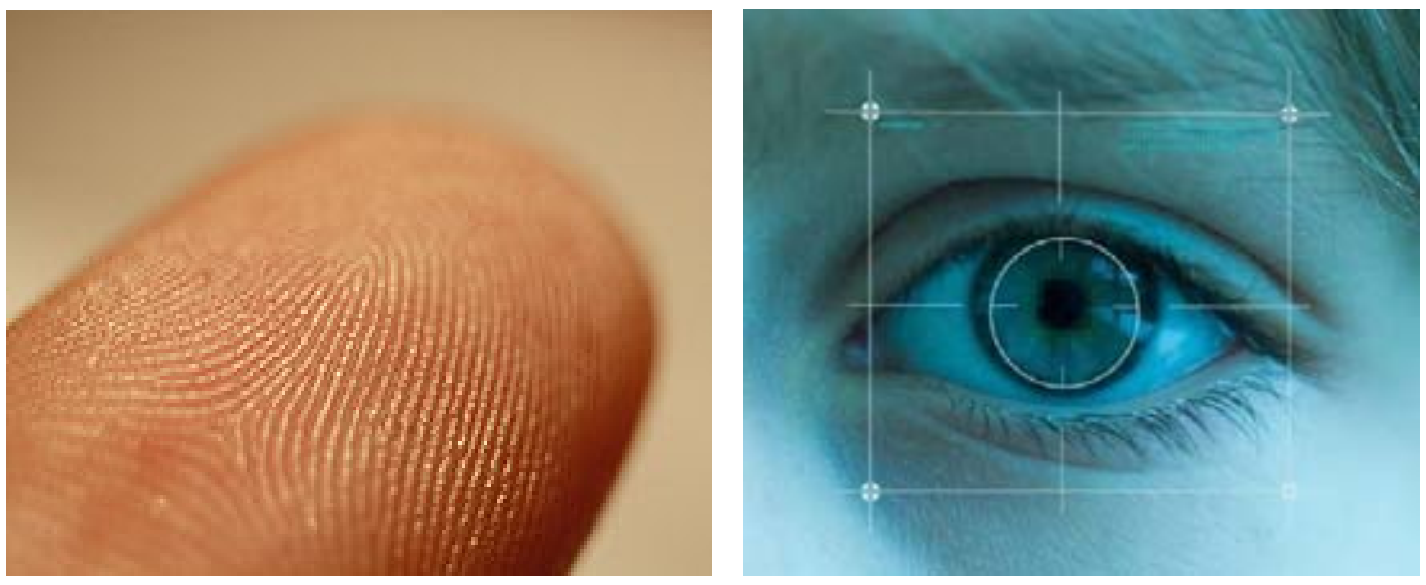

**Fig. 8.** A fingerprint and iris

Among the above symmetric identities, a watermark, trademark, and holographic label are collective identities of which each is assigned the same kind of object, a two-dimensional code, electronic supervision code, fingerprint, and iris are individual identities, and a stamp and password may be either collective identities or individual identities.

Note that because the symmetric identities are easily copied, imitated, forged, or liable to being manipulated by an inside jobber, they are insecure sometimes [25].

### 2.2.2 Asymmetric Identities

***Definition 5***: A cast or a string which is verified through the matching the left half with the right half or the public key with the private key is called an asymmetric identity [26].

The evolution of asymmetric identities also experiences two phases.

◈ Classical Asymmetric Identity

It is only a BingFu (namely a commander's tally) which is separated into two asymmetric halves, and appeared in China at least two thousand and two hundred years ago (Fig. 1 & Fig. 2) [13].

◈ Modern Asymmetric Identities

They include a RFID (Radio Frequency IDentity) which is embedded with a chip, occurred roughly in 1983 [27], and adopted digital signatures for anti-counterfeit in the late 1990's (Fig. 9) [28] [29], and a BFID (BingFu Identity Digitized) which is without a chip, originated from the early 2012's (Fig. 10) [30], and is composed of 16-22 printable characters corresponding to a digital signature.

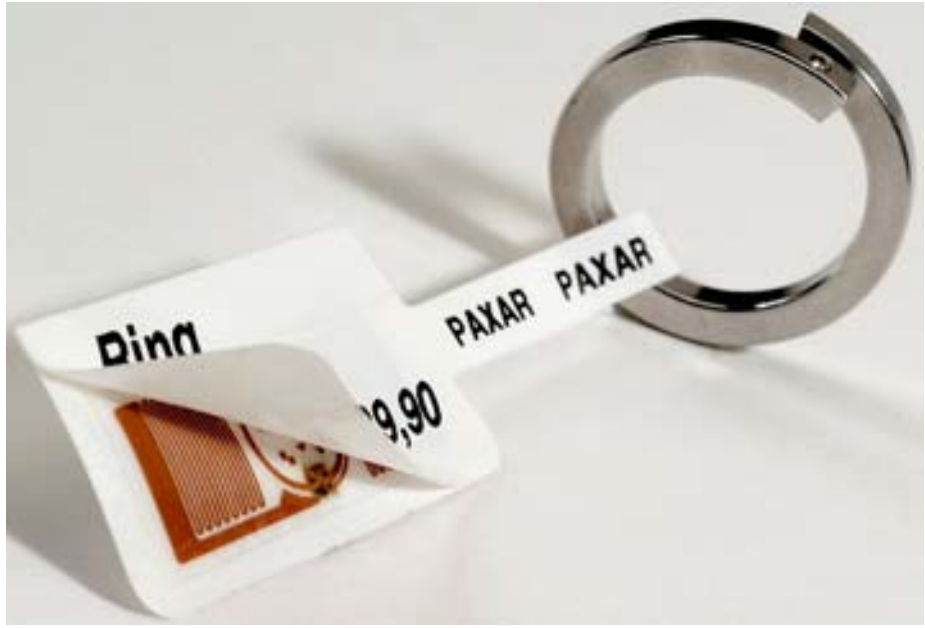

**Fig. 9.** A RFID with a chip storing a digital signature

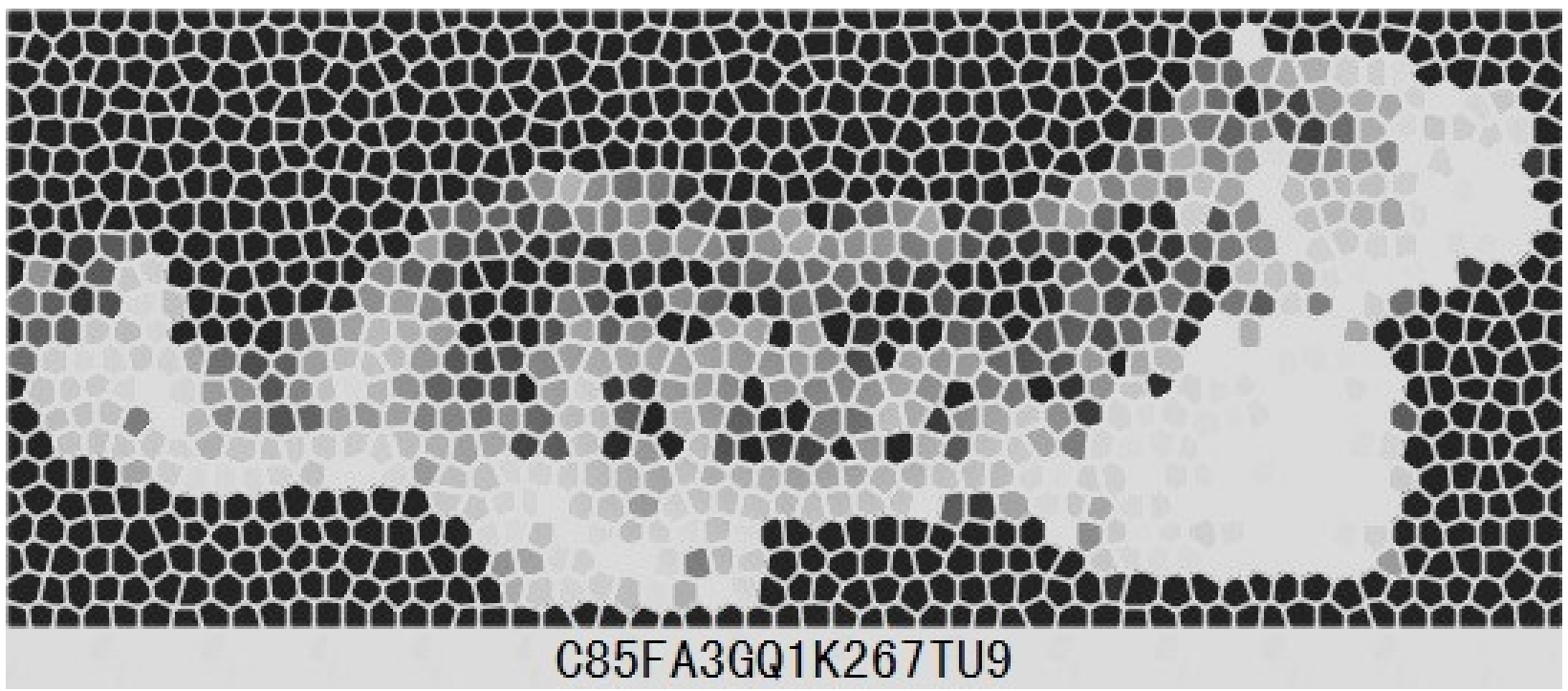

**Fig. 10.** A BFID consisting of 80 bits equivalent to 16 characters

From now on, an asymmetric identity just indicates a modern asymmetric identity for a classical asymmetric identity is not applied any longer.

According to Definition 5, an asymmetric identity has the following four properties [26]:

① Uniqueness: An asymmetric identity derives from a digital signature hiding the distinctive information of an object and being actually different from one another. Thus, it can uniquely represent one object in a domain.

② Anti-forgery: An asymmetric identity is obtained through a digital signing algorithm. Only can a specific public key which corresponds to the private key of a subject check it, and other public keys cannot check it. Consequently, the sufficiency of anti-forgery can be ensured.

③ Hiddenness: An asymmetric identity contains obscurely the characteristic of an object — a serial number, scaled time, biological attribute for example, and the private key of a subject.

④ Asymmetry: An asymmetric identity is confected with a private key, verified with a public key, and different from a symmetric identity which has no key or only one key.

Summarily, an asymmetric identity may produce important effects on authentication, anti-forgery, source tracing, and distribution monitoring.

***Definition 6***: An asymmetric identity which consists of only 16-22 characters, and does not need to be stored in a chip is called a lightweight asymmetric identity.

Because an asymmetric identity confected through the optimized REESSE1+ signing scheme may be 16-22 characters long (see Section 3.1), it is lightweight in terms of Definition 6 [26]. A lightweight asymmetric identity is also called a BFID.

#### 2.2.3 Conversion of a Symmetric Identity to an Asymmetric Identity

In remote service mode based on the Internet, the employment of a symmetric identity is probably insecure. We analyze an affair that a fingerprint as a password is used to log on a remote e-mail server. Under the circumstances, there exist two security risks:

① a fingerprint template stored in the server may be acquired by inside unauthorized technicians;

② a fingerprint instance transmitted on the Internet may be wiretapped by deliberate eavesdropers on logon.

Therefore, when a symmetric identity such as a fingerprint or string password is used remotely, it needs to be converted into an asymmetric identity through a hash function and a lightweight digital signing algorithm by the owner of a fingerprint or a secret string who holds a private key and a public key. The latter is placed on a united verification platform or a server.

Although any digitized symmetric identity may be converted into an asymmetric identity, sometimes the mixed or combined utilization of a symmetric identity and an asymmetric identity is needed.

### 2.3 Comparison between a Symmetric Identity and an Asymmetric Identity

A symmetric identity and an asymmetric identity act as different roles. A comparison between them may be made according to the definitions and facts (Tab. 1).

| | verifying method | Bearing character of an object | Inside job | United verification | Anti-forgery | Source tracing | Internet of things based on clouds |
|---|---|---|---|---|---|---|---|
| Sym. ID | Symmetric | No | Possible | Infeasible | No | No | Infeasible |
| Asym. ID | Asymmetric | Yes | Impossible | Feasible | Yes | Yes | Feasible |

**Tab. 1.** A comparison between a sym. identity and an asym. identity

It is well known that a symmetric identity such as an official stamp or electronic supervision code is easily imitated through an inside job or outside job since it does not bear the characteristic of an object [31]. It follows that an official stamp or electronic supervision code attached to authority is usually considered as trustworthy, but not secure. Note that trustworthiness does not equal security.

We can understand from Tab. 1 that the two primary advantages of an asymmetric identity are the prevention of inside jobs and the realizability of a united verification platform.

Asymmetric identities do not repulse symmetric identities. On some occasions, they are employed miscellaneously.

### 2.4 Comparison between a Symmetric Identity and an Asymmetric Identity

A RFID may be used for the anti-counterfeit of a merchant article, and similarly, a BFID may be used for the anti-counterfeit of a merchant article, especially a consumable commodity. Why is a BFID brought in yet? This needs to make a comparison between them (Tab. 2).

| | Security | Basic algorithm | Character-length | Storage | Sensing | Cost | Con-venience | Environ-mentalism |
|---|---|---|---|---|---|---|---|---|
| RFID | 2 ^ 80 | ECC\160 or RSA\1024 | 48 or 204 | Chip | Radio frequency | High | Low | Weak |
| BFID | 2 ^ 80 | REESSE1+\80 | 16 | Scrip or others | Light ray or others | Low | High | Strong |

**Tab. 2.** A comparison between a RFID and a BFID

Note that there is 16 = 80 / 5, namely a character is twice-hexadecimal.

It can be understanded from Tab. 2 that a BFID is tightly pertinent to a new specific algorithm, lightweight, economical, green, and easy to use (no need of card readers — for example).

## 3 Foundation of Lightweight Asymmetric Identities

A digital signature in a RFID chip is generally produced with the ECC or RSA signing scheme [32][33][34] while a digital signature corresponding to a BFID is produced with the optimized REESSE1+ signing scheme [35][36]. It is well known that a digital signing scheme is devised on the basis of intractable computational problems which are primarily constructed on the subbasis of computational complexity theory [37][38].

Concretely speaking, the security of ECC is based on an elliptic curve discrete logarithm problem (ECDLP) [33][39], the security of RSA is based on an integer factorization problem (IFP) [34][40], and the security of REESSE1+, including a hash function resistant to birthday attack [41], is based on the three new hardnesses and one classical hardness: a multivariate permutation problem (MPP), an anomalous subset product problem (ASPP), a transcendental logarithm problem (TLP) [35][42], and a polynomial root finding problem (PRFP) which is equivalent to the fact that a polynomial of high degree has only exponential time solutions at present [43][44]. Similar to PRFP, the hardnesses MPP, ASPP, and TLP have no subexponential time solutions so far [36][42].

### 3.1 Optimized REESSE1+ Signing Scheme

The optimized REESSE1+ digital signing scheme includes three algorithms for key-pair generation, digital signing, and signature authentication.

A digital signature is the output of the digital signing algorithm which takes a private key and an information digest as input [35].

The length of an optimized REESSE1+ modulus may only be 80 bits under the security of magnitude 2 ^ 80 while the length of an ECC modulus is 160 bits under the same security. The length of an optimized REESSE1+ signature only is 160 bits while the length of a ECC signature is at least 320 bits under the same security [33][35].

#### 3.1.1 Key Generation Algorithm

Assume that $đ$, $Đ$, $T$, $S$ are pairwise coprime integers, where $đ \in [5, 2^8]$, $T \geq 2^9$, $Đ \geq 2^{54}$ containing a prime $\geq 2^{52}$, and $\lceil \lg(đĐT) \rceil \geq 64$.

INPUT: a modulus length $m$ with $80 \leq m \leq 96$;
a sequence length $n$ with $80 \leq n \leq m \leq 96$;
a set $\Lambda = \{2, 3, \ldots, 863\}$.

S1: Produce appropriate parameters $đ$, $Đ$, $T$.
Produce the first $n/2$ natural primes $p_1, \ldots, p_{n/2}$.
Randomly produce a coprime sequence $\{A_1, \ldots, A_n\}$ with $A_i \in \Lambda$.
Randomly produce a set $\Omega = \{+/-5, +/-7, \ldots, +/-(2n + 3)\}$,
where every sign +/– means that "+" or "–" is selected.

S2: Find a prime $M$ making $\lceil \lg M \rceil = m$, $(đĐT) \mid \bar{M}$ and $\prod_{i=1}^{k} p_i^{e_i} \mid \bar{M}$,

where $k$, $e_i$ and $p_k$ meet $\prod_{i=1}^{k} e_i \approx 2^8$ and $p_k < p_{n/2}$.

Pick $S \in (1, \bar{\boldsymbol{M}})$ making $\gcd(S, \bar{\boldsymbol{M}}) = 1$ and $S^{-1} \,\%\, \bar{\boldsymbol{M}}$ small.

S3: Pick $W, \delta \in (1, \bar{\boldsymbol{M}})$ meeting $\gcd(W, \text{đ}\,\text{Đ}) > 1$, $\gcd(\delta, \bar{\boldsymbol{M}}) = 1$ and $\|\delta\| = \text{đ}\,\text{Đ}T$.

S4: Compute $\alpha \leftarrow \delta^{(\delta^{\bar{o}} + \delta W^{\bar{o}-1})T}$, $\beta \leftarrow \delta^{W^{\bar{o}}T}$, $\hbar \leftarrow (W \prod_{i=1}^{n} A_i)^{-\delta S} (\alpha \delta^{-1}) \,\%\, M$,

where $\bar{\boldsymbol{o}} \approx \bar{\boldsymbol{M}} / 2$ is a big prime.

S5: Randomly produce pairwise distinct $\ell(1), \ldots, \ell(n) \in \Omega$.

S6: Compute $C_i \leftarrow (A_i W^{\ell(i)})^{\delta} \,\%\, M$ for $i = 1, \ldots, n$.

OUTPUT: $(\{C_i\}, \alpha, \beta)$ regarded as a public key;

$(\{A_i\}, \{\ell(i)\}, W, \delta, \text{Đ}, \text{đ}, \hbar)$ as a private key;

$(\bar{o}, n, S, T, M)$ as being in common.

This algorithm is called by an identity key management algorithm.

### 3.1.2 Digital Signing Algorithm

Assume that *hash* is a one-way compression function — the Juna hash for example.

INPUT: a private key $(\{A_i\}, \{\ell(i)\}, W, \delta, \text{Đ}, \text{đ}, \hbar)$; a file or message $F$.

S1: Let $H \leftarrow hash(F)$, whose binary form is $b_1 \ldots b_n$.

S2: Set $\underline{k} \leftarrow \delta \sum_{i=1}^{n} b_i \ell(i) \,\%\, \bar{\boldsymbol{M}}$, $G_0 \leftarrow (\prod_{i=1}^{n} A_i^{\neg b_i})^{\delta} \,\%\, M$.

S3: $\forall \bar{a} \in (1, \bar{\boldsymbol{M}})$ making $(\text{đ}T) \nmid \bar{a}$ and $\text{đ} \nmid (WQ) \,\%\, \bar{\boldsymbol{M}}$,

where $Q = (\bar{a}\,\text{Đ} + WH)\delta^{-1} \,\%\, \bar{\boldsymbol{M}}$.

S4: Compute $R \leftarrow (Q(\delta \hbar)^{-1})^{S^{-1}} G_0^{-1}$, $\bar{U} \leftarrow (RW^{\underline{k} - \delta})^{Q} \,\%\, M$,

$\bar{\boldsymbol{g}} \leftarrow \delta^{\bar{a}\,\text{Đ}} \,\%\, M$, $\xi \leftarrow \sum_{i=0}^{\bar{o}-1} (\delta Q)^{\bar{o}-1-i} (HW)^i \,\%\, \bar{\boldsymbol{M}}$.

S5: $\forall\, r \in [1, \text{đ}\, 2^{16}]$ making $\text{đ} \nmid (r U S + \xi) \,\%\, \bar{\boldsymbol{M}}$,

where $U = \bar{U} \bar{\boldsymbol{g}}^{r} \,\%\, M$.

S6: If $\text{đ} \nmid ((WQ)^{\bar{o}-1} + \xi + r U S) \,\%\, \bar{\boldsymbol{M}}$ then go to S5 else end.

OUTPUT: $(Q, U)$, a signature on the file $F$.

This algorithm is called by an identity confection algorithm.

### 3.1.3 Signature Authentication Algorithm

Assume that *hash* is the above one-way compression function.

INPUT: a public key $(\{C_i\}, \alpha, \beta)$; a file or message $F$; a signature $(Q, U)$.

S1: Let $H \leftarrow hash(F)$, whose binary form is $b_1 \ldots b_n$.

S2: Compute $\bar{G}_1 \leftarrow \prod_{i=1}^{n} C_i^{b_i} \,\%\, M$.

S3: Compute $X \leftarrow (\alpha Q^{-1})^{QUT} \alpha^{Q^{\bar{o}}} \,\%\, M$,

$Y \leftarrow (\bar{G}_1^{Q} U^{-1})^{UST} \beta^{HQ^{\bar{o}-1} + H^{\bar{o}}} \,\%\, M$.

S4: If $X = Y$ then the signature is valid and $F$ intact else the signature is invalid or $F$ modified.

OUTPUT: "yes" or "no".

This algorithm is called by an identity verification algorithm.

## 3.2 Juna Hash Function

***Definition 7***: Let $b_1 \ldots b_n \neq 0$ be a bit string. Then $\underset{\cdot}{b}_i$ with $i \in [1, n]$ is called a bit shadow if it comes from such a rule [41]:

① $\underset{\cdot}{b}_i = 0$ if $b_i = 0$;

② $\underset{\cdot}{b}_i = 1 +$ the number of successive 0-bits before $b_i$ if $b_i = 1$; or

③ $\underset{\cdot}{b}_i = 1 +$ the number of successive 0-bits before $b_i$ + the number of successive 0-bits after the rightmost 1-bit if $b_i$ is the leftmost 1-bit.

For example, let $b_1 \ldots b_8 = 01010100$, then $\underset{\cdot}{b}_1 \ldots \underset{\cdot}{b}_8 = 04020200$.

***Definition 8***: Let $\underset{\cdot}{b}_1 \ldots \underset{\cdot}{b}_n$ be the bit shadow string of $b_1 \ldots b_n \neq 0$. Then $\bar{b}_i = \underset{\cdot}{b}_i 2^{\partial_i}$ with $i \in [1, n]$ is called a bit long-shadow, where $\partial_i = b_{i + (-1)^{\lfloor 2(i-1)/n \rfloor}(n/2)} = 0$ or $1$ [41].

For example, let $b_1 \ldots b_8 = 01010100$, then $\bar{b}_1 \ldots \bar{b}_8 = 08020400$.

The Chaum-Heijst-Pfitzmann hash function, a non-iterative one based on a DLP, is appreciable [45].

The new non-iterative hash function is constituted of two algorithms which contain two main parameters $m$ and $n$, where $m$ denotes the bit-length of a modulus utilized in the new hash, $n$ denotes the bit-length of a short message or a message digest from a classical hash function, and there are $80 \le m \le 232$ with $80 \le m \le n \le 4096$.

Additionally, $\Lambda$ and $\Omega$ are two integral sets. Their lengths are selected as $2^{10} \le |\Lambda| \le 2^{32}$ and $n \le |\Omega| = \tilde{n} \le 2^{32}$, and moreover make $2n^5|\Omega||\Lambda|^5 \ge 2^m$. Notice that $2^{10} \le |\Lambda| \le 2^{32}$ means $10 \le \lceil \lg \bar{P} \rceil \le 32$.

For example, as $m = 80 \le n$, there should be $|\Lambda| = 2^{10}$ and $|\Omega| = n$; as $m = 96 \le n$, should $|\Lambda| = 2^{12}$ and $|\Omega| = n$; as $m = 112 \le n$, should $|\Lambda| = 2^{14}$ and $|\Omega| = n$; as $m = 128 \le n$, should $|\Lambda| = 2^{16}$ and $|\Omega| = 2^{12}$; as $m = 232 \le n$, should $|\Lambda| = 2^{32}$ and $|\Omega| = 2^{32}$.

### 3.2.1 Initialization Algorithm

This algorithm is employed by an authoritative third party or the owner of a key pair, and only needs to be executed one time.

INPUT: the bit-length $m$ of a modulus with $80 \le m \le 232$;
the item-length $n$ of a sequence with $80 \le m \le n \le 4096$;
the maximal prime $\bar{P}$ with $10 \le \lceil \lg \bar{P} \rceil \le 32$;
the size $\tilde{n}$ of the set $\Omega$ with $2\tilde{n}n^5\bar{P}^5 \ge 2^m$ and $n \le \tilde{n} \le 2^{32}$.

S1: Produce $\Lambda \leftarrow \{2, 3, \ldots, \bar{P}\}$;
produce a random coprime sequence $\{A_1, \ldots, A_n \mid A_i \in \Lambda\}$.

S2: Find a prime $M$ with $\lceil \lg M \rceil = m$ such that $\bar{M} / 2$ is a prime,
or the least prime factor of $\bar{M} / 2$ is bigger than $4n(2\tilde{n} + 3)$.

S3: Pick $W \in (1, \bar{M})$ making $\|W\| \ge 2^{m - \lceil \lg \bar{P} \rceil}$;
pick $\delta \in (1, \bar{M})$ making $\gcd(\delta, \bar{M}) = 1$.

S4: Randomly yield $\Omega \leftarrow \{+/-5, +/-7, \ldots, +/-(2\tilde{n} + 3)\}$;
randomly select pairwise distinct $\ell(i) \in \Omega$ for $i = 1, \ldots, n$.

S5: Compute $C_i \leftarrow (A_i W^{\ell(i)})^{\delta} \% M$ for $i = 1, \ldots, n$.

OUTPUT: an initial value $(\{C_i\}, M)$ which is public to the people.

A private parameter $(\{A_i\}, \{\ell(i)\}, W, \delta)$ may be discarded, but must not be divulged.

At S3, to seek $W$, let $W \equiv g^{\bar{M}/F}$ ($\% M$), where $g$ is a generator of $(\mathbb{Z}_M^*, \cdot)$ obtained through Algorithm 4.80 in Section 4.6 of [46], and $F < 2^{\lceil \lg \bar{P} \rceil}$ is a factor of $\bar{M}$.

At S4, $\Omega = \{+/-5, +/-7, \ldots, +/-(2\tilde{n} + 3)\}$ indicates that $\Omega$ is one of $2^{\tilde{n}}$ potential sets, indeterminate, and unknown to the public, where "+/–" means the selection of the "+" or "–" sign. Notice that in the arithmetic modulo $\bar{M}$, $-x$ represents $\bar{M} - x$.

### 3.2.2 Compression Algorithm

This algorithm is employed by one who wants to obtain a short message digest.

INPUT: an initial value $(\{C_1, \ldots, C_n\}, M)$, where $\lceil \lg M \rceil = m$ with $80 \le m \le n \le 4096$;
A short message (or a digest from a classical hash function) $b_1 \ldots b_n \ne 0$.

S1: Set $k \leftarrow 0$, $i \leftarrow 1$.

S2: If $b_i = 0$ then
S2.1: let $k \leftarrow k + 1$, $\underset{\cdot}{b}_i \leftarrow 0$
else
S2.2: if $i = k + 1$ then let $\bar{s} \leftarrow i$;
S2.3: let $\underset{\cdot}{b}_i \leftarrow k + 1$, $k \leftarrow 0$.

S3: Let $i \leftarrow i + 1$;
if $i \le n$ then go to S2.

S4: Compute $\underset{\cdot}{b}_{\bar{s}} \leftarrow \underset{\cdot}{b}_{\bar{s}} + k$.

S5: Compute $\underline{d} \leftarrow \prod_{i=1}^{n} C_i^{\bar{b}_i} \% M$,

where $\bar{b}_i = \underset{\cdot}{b}_i 2^{\partial_i}$ with $\partial_i = b_{i + (-1)^{\lfloor 2(i-1)/n \rfloor}(n/2)}$.

OUTPUT: a digest $\underline{d} \equiv \prod_{i=1}^{n} C_i^{\bar{b}_i}$ (% $M$) of which the bit-length is $m$.

It is easily known from Definition 8 that the max of $\{\bar{b}_1, \ldots, \bar{b}_n\}$ is less than or equal to $n$ when $b_1 \ldots b_n \neq 0$.

# 4 United Verification Platform for BFIDs and Identification Systems

## 4.1 Topological Structure of the Platform

The platform working in cloud computing mode and being reached through a domain name or a mobile number supplies shared storage and verification services for a wide variety of users or lessees (Fig. 11). For security, the cloudland is partitioned into an inner network and an outer network.

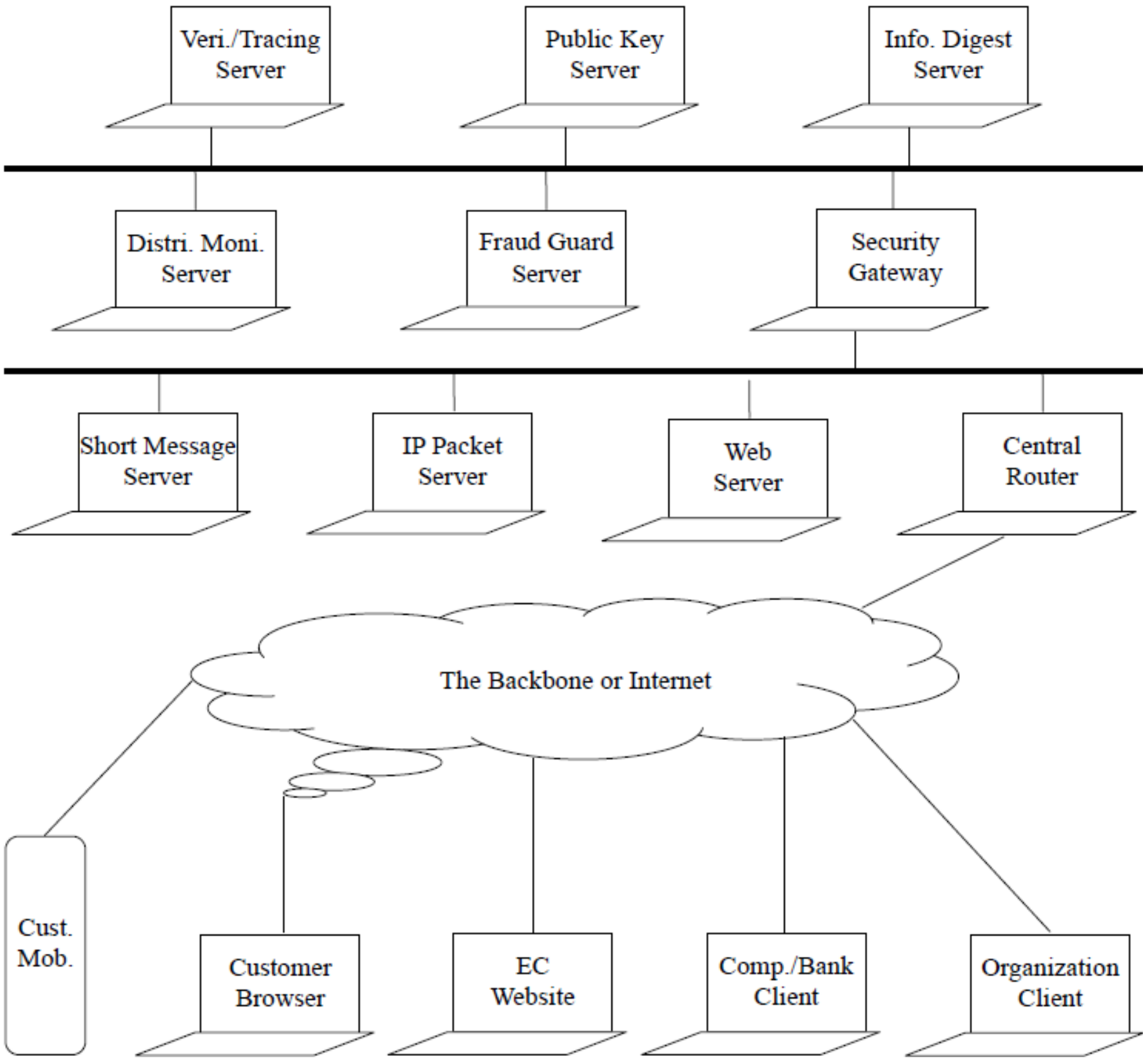

**Fig. 11.** Topological structure of the united verification platform for BFIDs

Between the inner network and the outer network, there is a security gateway which can prohibit illegal IP packets from invading the inner network. Additionally, the platform will have a strict identification mechanism which can prevent unauthorized persons from accessing and modifying databases. The platform does not need strict encryption measures since all data on it are allowed being in a public state.

The platform is composed of a verification server, source tracing server, public key server, information digest server, distribution monitoring server, fraud guard server, short message server, IP packet server, web server, etc.

• Verification / Source Tracing Server

Being responsible for the checking computation of BFIDs, and returning the result and relevant information to an inquirer.

• Public Key Server

Storing all public keys of enterprises, organizations, institutions, and persons. These public keys may be accessed, but may not be modified.

• Information & Digest Server
Storing characteristic information, description data, source information, or their digests of persons and things (commodities, bills, notes, diplomas, certificates, websites, programs, etc).

• Distribution Monitoring Server
Supervising and managing the warehouse-outing, delivery, passage, and marketing of packaged merchandise articles so as to prevent the overlap of sale regions and guarantee the retrospection of sale channels.

• Fraud Guard Server
Detecting malicious programs, fishing websites, fabricated IP addresses, counterfeited commodities (especially food or drug articles), faked bills or notes, forged diplomas or certificates, imitated credentials or licenses on the basis of BFID verification results, as well as reporting occurrences to related enterprises, institutions, or administrations who take charge of disposing frauds.

• Short Message Server
Being responsible for receiving a BFID sent by a customer or consumer with a mobile, and directly returning a verification result and a piece of source information to the customer or consumer.

• IP Packet Server
Being responsible for receiving a program BFID from a computer outside the platform, and directly returning a verification result in the form of a IP packet to the computer.

• Web Server
Providing web pages for question inquiry and BFID checking service, and return an answer or verification result to the browser of a customer or user.

### 4.2 Software Composition of an Identification System

Multiple identification systems may be developed on the basis of the united verification platform. Each of these systems consists of a server terminal, client terminal, and inquiry terminal.

A server terminal shares the united verification platform, a client terminal needs to be developed individually according to the business of a different client — an enterprise or institution for example, and an inquiry terminal also needs to be developed individually according to the medium of a different inquirer a user or customer for example.

• Software at a Server Terminal
Including a BFID verifier, a source tracing subsystem, a distribution monitoring subsystem, a fraud guarder, a short message exchanger, an IP packet exchanger, a web page exchanger, a user register, etc.

• General Software at a Client Terminal
Involving an asymmetric foundational operator, a one-way hash function, a key-pair generator, a key-pair keeper, a digital signer, etc.

• Individual Software at a Client Terminal
Involving a public key cloudland storage subsystem, an identity confection subsystem, a source information transfer subsystem, a label/mark concocting subsystem, a multiple identities handling subsystem, etc.

• Software at an Inquiry Terminal
Including identity information sensor, identity validity query subsystem, etc.

## 5 Applications of Idology

Obviously, idology has its applicative value in public security and network security.

### 5.1 Applications in Public Security

Fraud in the real world is rampant, which damages greatly the development of national economy, hence, it should be prevented at technical level.

• Banknote Anti-forgery System Based on BFIDs

The central bank gives every banknote a BFID which is visibly printed on the surface of the banknote without a chip, and hides the private key of the central bank and the characteristic of the banknote — the currency number, par value, issuing bank name, issuing date, currency version etc for example. The BFID sent with the mobile of a customer is checked by the united verification platform, and the platform can detect whether a BFID is imitated, and a banknote is forged through analyzing big BFID data [47].

• Diploma Anti-forgery System Based on BFIDs

A college or university grants every graduate a diploma on which a BFID is visibly printed. The BFID conceals the private key of the college or university and the characteristic of the graduate — the name, birthday, resident ID number / social insurance number, graduating time, graduating university, specialty etc for example. Because the ID number or social insurance number of the graduate is unique and inimitable, the BFID is also inimitable.

When the diploma of a graduate is examined by an employer, a related BFID is sent to the united verification platform alongside of the ID card or insurance card of the graduate being examined [48].

• Food/Drug Anti-counterfeit and Source Tracing System Based on BFIDs

A food (or drug) mill assigns to every food (or drug) article a BFID which is printed on the label of the food (or drug) article, covered with a layer of luminescent powder, and hides the private key of the mill and the characteristic or source information of the food (or drug) article — the product number, material component, production date, expiration date, mill name, mill address etc for example.

The coat on a BFID needs to be scraped off when it is inputted or scanned to a mobile for attestation. The united verification platform will return the "True" value, source information, and marketing channel if the BFID passes the examination [26].

• Passport Anti-forgery System Oriented to Antiterrorism

Most of terrorists utilized fake passports to sneak into intendedly attacked countries (Fig. 12) [49].

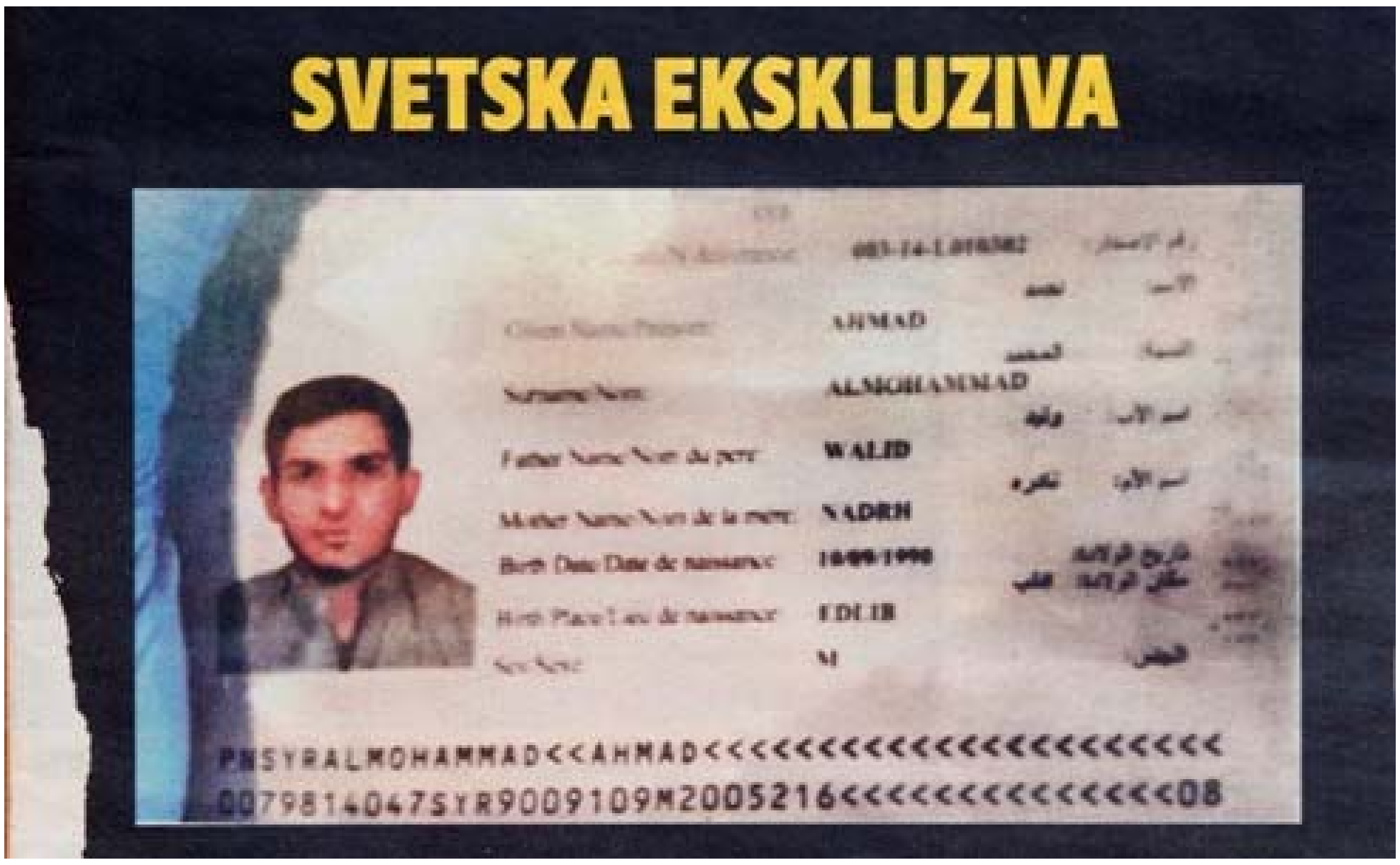

**Fig. 12.** Serbian magazine Blic displays a fake Syrian passport found at a scene of the Paris attacks

Therefore, to prevent passports from being forged, should let a passport contain a lightweight asymmetric identity such as a BFID and a biological characteristic. The former hides the private key of an issuing government and the personal material of a related holder, and ensures the coherence between the issuing government and the passport; the latter is the head-photo, fingerprint, or iris of the related holder which ensures the coherence between the passport and the related holder. By transmissibility, there is the coherence between the issuing government and the related holder.

The lightweight asymmetric identity should be visibly printed on a page of the passport so as to be capable of being checked in manual way through a united verification platform. Clearly, some protocols on and standards for lightweight asymmetric identities and the united verification platform should be proposed to a pertinent international association and sufficiently discussed among most nations before the final versions are concluded.

## 5.2 Applications in Network Security

The Internet is innately imperfect, which leaves occasions to cheaters.

• Prevention of Fictional IP Addresses

IETF published Internet Protocol version 6 (shortly IPv6) in 1998 [50]. IPv6 adopts a 128-bit address format allowing 2128 addresses, which means that almost every device on the Internet may be assigned an IP address for identification and location definition. However, it has two prominent flaws:

① a nation identifier is not designated in the global routing prefix of an IP address;

② there is no mechanism for precluding fictional IP addresses or IP addresses fraud although there is the Internet Protocol Security (IPsec) attached to IPv6 which is developed only to prevent the content of an IP packet from being divulged as well as the source and destination addresses in an IP packet from being tampered.

Hence, IPv6+ is suggested by the authors. The dominant difference between IPv6+ and IPv6 is the partition of a source or destination address and the prevention of fictional IP addresses. In IPv6+, the structure of a source or destination address is as follows (Fig. 13):

| 8 bits | 24-32 bits | 8-16 bits | 80 bits |
|---|---|---|---|
| Nation ID | Domestic Routing Indicator | Subnet ID | Interface ID |

**Fig. 13.** Structure of an IPv6+ Address

In an IPv6+ address, the nation ID which is an 8-bit number represents the boundary of a national territory in the cyberspace, needs to be uniformly assigned among major countries, and should be coincident with the international calling code of a country — 86 being of China for example.

The domestic routing indicator which is a 24-32 bit number represents the approach to a province (state), municipality (county), city, or town.

The subnet ID which is an 8-16 bit number represents a subsystem of interconnections within a system, and it allows the components to communicate directly with each other.

The interface ID equivalent to a BFID which is an 80 bit number represents a host, router, or other device on a subnet, and hides the private key of a national cyberspace administration and the characteristic of an interface such as the domain name (globally unique), EUI-64 address (also globally unique), nation ID, domestic routing indicator, subnet ID, etc.

When a destination host receives a IP packet, it will extract a source address from the IP packet, and send the source address to the united verification platform which will check the validity of the source address with the public key of the national cyberspace administration, and return a result to the host. Note that a IP packet should be simultaneously signed with the private key of a user on a related source host.

• Dynamic Password Equivalent to a BFID

A dynamic password of which the checking is not a direct comparison or hash comparison is confected through the optimized REESSE1+ signing scheme, hides the private key source or biologic fingerprint of a user and the characteristic of a login occasion such as the user name, login date, login time, machine name, etc, is thoroughly different on every login, and is checked with the user's public key that is fetched to a server terminal in advance. Obviously, a dynamic password is substantially equivalent to a BFID which is so short as to be capable of being checked fast.

The advantage over a classical password of a dynamic password is that it may protect a confidential fingerprint or private key source against being exposed to an eavesdropper in a transmission process or to an inside job worker at a server terminal.

• Official Document against Forgery and Tampering

When an official document of an organization or administration is promulgated on the Internet, it should bear a BFID which is so short as to be capable of being placed at the document's file name position [35][41]. The BFID conceals the private key of the organization or administration and the characteristic of the document — the title, script number, key words, promulgator name, promulgating date, etc for example, is checked by the united verification platform, and can defend the document against being forged or tampered.

• Prevent Computer Viruses

A program is given an asymmetric identity BFID by its developer, and the BFID is put in the program's file name [35][41]. When the program is executed, the BFID will be captured, sent to the united verification platform, and verified with the public key of the developer. If the BFID is ineligible, the execution of the program will be terminated. It is impossible that a computer virus acquire an eligible BFID, and this way, computer viruses may be prevented.

• Attestation by a Real Name against Privacy Divulgence in Cyberspace

When a citizen registers at a server with his own real identity, he needs to input from a keyboard his account-name, BFID (equivalent to a real name), profession, affiliation, etc, where the BFID will be sent to the united verification platform storing a variety of public keys for examination. Note that a traditional password is not necessary for registration due to a dynamic password being available.

Such a BFID as a network identity of a citizen hides the private key of a national population administration and the characteristic of a resident such as the legal identity — the Chinese ID number for example, legal name, gender, home address, telephone number etc. Note that the Chinese ID number despite being nationally unique is not allowed to be inputted into a server on the registration of a citizen because it contains the privacy birth date of the citizen.

## 6 Conclusion

Omnipresent fraud in both the real world and the cyberspace causes the international society giant economic losses [1][2][3][4]. The prevention of fraud needs not only behavioral legislation but also technical innovation. The latter is more urgent.

The logic makes it clear that the prevention of fraud is essentially the identification of a person or thing in the real world and the cyberspace.

Facts elucidate that ubiquitous old-line symmetric identities are incompetent for anti-fraud, and asymmetric identities represent a new direction in anti-fraud.

Extensive and favorable application demands leads the naissance of a new discipline — idology which is the study of knowledge on identifications of persons or things in brief.

The aim of idology is the prevention of fraud while the aim of cryptology is the prevention of secret-outing, and thus idology should be independently considered and treated. If they are confused, then the fulfillment of anti-fraud tasks all over the world will be affected severely.

Along with the coming of quantum computer era, researches on post-quantum idology which is resistant to quantum computation attack namely Shor algorithm attack will be made [51].

Many problems relevant to idology are faced with us, and need to be seriously resolved.

## Acknowledgment

The authors would like to thank the Acad. Jiren Cai, Acad. Zhongyi Zhou, Acad. Changxiang Shen, Acad. Zhengyao Wei, Acad. Binxing Fang, Acad. Guangnan Ni, Acad. Andrew C. Yao, Acad. Jinpeng Huai, Acad. Wen Gao, Prof. Jie Wang, Rese. Hanliang Xu, Rese. Dengguo Feng, Rese. Dali Liu, Rese. Qiquan Guo, Rese. Rui Yu, Prof. Zhiying Wang, Prof. Ron Rivest,

Prof. Moti Yung, Prof. Dingzhu Du, Acad. Xiangke Liao, Acad. Wenhua Ding, Prof. Huaimin Wang, Prof. Jianfeng Ma, Prof. Heyan Huang, Prof. Zhong Chen, Prof. Jiwu Jing, Prof. Gongxuan Zhang, Prof. Yixian Yang, Prof. Maozhi Xu, Prof. Bing Chen, Prof. Xuejia Lai, Prof. Yongfei Han, Prof. Yupu Hu, Prof. Ping Luo, Acad. Wei Li, Acad. Xicheng Lu, Prof. Dingyi Pei, Prof. Huanguo Zhang, Prof. Mulan Liu, Prof. Bogang Lin, Prof. Renji Tao, Prof. Quanyuan Wu, and Prof. Zhichang Qi for their important suggestions, corrections, and helps.

This work is supported by MOST with Project 2007CB311100 and 2009AA01Z441. Corresponding email: reesse@126.com.